\documentclass[aps,prb,preprint,groupedaddress,showpacs,showkeys]{revtex4}
\usepackage{graphicx}
\usepackage{amsbsy}

\begin{document}

\title{Stochastic linear scaling for metals and non metals}
\author{Florian R. Krajewski}
\affiliation{Computational Science, Department of Chemistry and Applied Biosciences, ETH
Zurich, USI Campus, Via Giuseppe Buffi 13, CH-6900 Lugano, Switzerland}
\author{Michele Parrinello}
\affiliation{Computational Science, Department of Chemistry and Applied Biosciences, ETH
Zurich, USI Campus, Via Giuseppe Buffi 13, CH-6900 Lugano, Switzerland}
\date{\today}

\begin{abstract}
Total energy electronic structure calculations, 
based on density functional theory or on the 
more empirical tight binding approach, are generally 
believed to scale as the cube of the number of electrons. 
By using the localisaton property of the high temperature density 
matrix we present exact deterministic 
algorithms that scale linearly in one dimension and 
quadratically in all others. We also introduce a 
stochastic algorithm that scales linearly with system size. 
These results hold for metallic and non metallic systems 
and are substantiated by numerical calculations on model systems.
\end{abstract}

\maketitle










Total energy electronic structure calculations and molecular dynamics
simulations based on density functional theory (DFT) or on the tight binding
approach have been very successful in describing a large variety of phenomena
and are finding increasing application in many fields of science. However
even with present-day computer technology the size of the systems that can
be studied is very restricted. This is due to the cubic dependence of the
commonly used algorithms on the number of particles. This makes a daunting
task of calculating the electronic properties of systems as large as those
that are for instance of interest to nanotechnology and biochemistry.

Some 15 years ago it was realized that this need not be so and that
linearly scaling algorithms could be devised\cite{yang91,gall92}.
Up to now large number of linearly scaling  algorithms have been proposed\cite{goe99}, 
but they are not devoid of problems and for a variety of reasons
they are not yet routinely used. Most algorithms rely on the fact that the
wavefunctions can be  localized and have an exponential decay leading to a
sparse  Hamiltonian. This property does not hold when the gap between
occupied and unoccupied levels vanishes, as in the case of metals for which
it has  proven difficult to obtain linearly scaling algorithms. 

Here we propose a new approach to this problem that does not rely on an
ability to localize the wavefunctions and is therefore equally applicable
to metallic and non-metallic systems. We introduce a series of algorithms
which defy the commonly accepted wisdom that DFT calculations are of $%
O\left( N^{3}\right) $. In fact they are of $O\left( N^{2}\right) $ and even
of $O\left( N\right) $. Furthermore we propose a stochastic algorithm that
is linear scaling in all dimensions. A feature which sets our method apart
from others is that it scales with the volume of the system and not with the
number of electrons. A bonus, if one treats atomic species that are rich in
electrons.

We work at finite temperature $\frac{1}{\beta }$ in the grand canonical
ensemble, where the number of particles is controlled by the chemical
potential $\mu $ and the relevant thermodynamic potential for spinless
fermions is \cite{mer65}: 
\begin{equation}
\Omega _{f}=-\frac{1}{\beta }\ln \det \left( {1}+e^{\beta \left( \mu -{H}%
\right) }\right)   \label{eq:gp}
\end{equation}%
where $H$ is a single particle Hamiltonian $H=-\frac{1}{2}\nabla +V\left(
r\right) $. In the $\beta \rightarrow \infty $ limit%
\begin{equation}
\Omega _{f}=\sum_{i}\varepsilon _{i}-\mu N  \label{eq:ztl}
\end{equation}%
where the sum runs over the $N$ lowest occupied states and $N$ is the total
number of particles. Limiting oneself to the consideration of free electrons
in an external field is not as restrictive as it might at first seem. In
fact it is well known that the full density functional can be obtained from
the sum over the occupied orbital energy in Eq.~(\ref{eq:ztl}), if the
external potential of the Hamiltonian $H$ is the self-consistent DFT
potential and double counting terms are subtracted\cite{parr}. Since the
latter can be calculated with linear effort, finding a method for evaluating 
$\Omega _{f}$ in $O(N)$ operations solves the algorithmically hard part of the
problem.

We make use of the identity: 
\begin{equation}
\left( {1}+e^{\beta \left( \mu -{H}\right) }\right) =\prod_{l=1}^{P}\left( {1%
}+e^{i\frac{\pi }{P}\left( 2l-1\right) }\;e^{\frac{\beta }{P}\left( \mu -{H}%
\right) }\right)   \label{eq:prod}
\end{equation}
which is valid for any even $P$ and the product goes over all the complex $%
P^{\mathrm{th}}$ roots of $-1$. Using this decomposition one finds 
\begin{equation}
\Omega _{f}=-\frac{1}{\beta }\sum_{l=1}^{P}\ln \det \left( {\boldsymbol M}%
(l)\right)   \label{eq:om}
\end{equation}
with ${\boldsymbol M}(l)={1}+e^{i\frac{\pi }{P}\left( 2l-1\right) }\;e^{%
\frac{\beta }{P}\left( \mu -{H}\right) }$. This is rather more complicated
than Eq.~(\ref{eq:gp}) but has the advantage that it involves the propagator 
$e^{\frac{\beta }{P}\left( \mu -{H}\right) }$ rather than the more difficult 
$e^{\beta \left( \mu -{H}\right) }$. In fact if $P$ is large enough one can
use for $e^{\frac{\beta }{P}\left( \mu -{H}\right) }$ one of the many 
high-temperature representations of the exponential operator, such as the one
based on the expansion in terms of Chebychev polynomials\cite{goe94,goe95} or
the Trotter decomposition\cite{chand81}, as is commonly done in the numerical
evaluation of path integrals. In this first implementation we shall employ the
latter, which uses the identity:
\begin{eqnarray}
&&
\left \langle r \left \vert 
      e^{ \frac{\beta }{P} \left( \mu -{H}\right) }
\right\vert r^{\prime }\right\rangle
\\
&=&e^{\frac{\beta }{2P}\left( \mu-V\left( r\right) \right) }
  e^{-\frac{mP}{2\hbar ^{2}\beta }\left(r-r^{\prime }\right) ^{2}}
  e^{\frac{\beta }{2P}\left( \mu -V\left( r^{\prime}\right) \right) } 
\nonumber
\end{eqnarray}
valid to $O\left( \left( \frac{\beta }{P}\right) ^{3}\right) $. For large $P$
$\ \ \ e^{-\frac{mP}{2\hbar ^{2}\beta }\left( r-r^{\prime }\right) ^{2}}$
decays very rapidly and most elements of $e^{\frac{\beta }{P}(\mu -{H})}$
can be neglected, leading to a sparse matrix which is a key factor for
obtaining a linear scaling algorithm. Since this property does not depend on
the possibility of localizing the wavefunctions this makes our method
suitable also for metals. Alavi and Frenkel\cite{alavi92} have shown that 
the high temperature density matrix in a cubic lattice of spacing $\delta $
can equally well be written as: 
\begin{eqnarray}
&&\left\langle i\left\vert e^{\frac{\beta }{P}\left( \mu -H\right)
}\right\vert j\right\rangle  \\
&=&C\left\{ 
\begin{array}{ll}
e^{\frac{\beta }{P}\left( \mu -V\left( i\right) \right) } & i=j \\ 
e^{\frac{\beta }{2P}\left( \mu -V\left( i\right) \right) }e^{-\frac{mP\delta
^{2}}{2\hbar ^{2}\beta }}e^{\frac{\beta }{2P}\left( \mu -V\left( j\right)
\right) } & \!\!%
\begin{array}{l}
\mathrm{if}\;i\;\mathrm{and}\;j\;\mathrm{are} \\ 
\mathrm{first\;\;neighbors}%
\end{array}
\\ 
0 & \mathrm{otherwise}%
\end{array}%
\right.   \nonumber
\end{eqnarray}%
where $i$ and $j$ are lattice site  indices, $P$ is related to the lattice
spacing by $P=3.432\hbar ^{2}\beta /(m\delta ^{2})$ and $C$ is a
normalization constant. This expression is compatible with the
approximations made so fare, leads to an elegant lattice model with nearest
neighbor interactions and at the same time does not make the problem any
less complex, nor does it alter its scaling behaviour.

Using this representation of the density matrix it is easily seen that in
one dimension the matrices ${\boldsymbol M}\left( l\right) $ that 
appear in Eq.~(\ref{eq:om}) 
are tridiagonal. Since the determinant of a tridiagonal matrice can be
computed in $O(M)$ operations, we arrive at the interesting result that
linear scaling is exact in 1d. Moving to higher dimensions the $M\left(
l\right) $ matrices become block tridiagonal where the dimension of each
block is $m=M^{\left( d-1\right) /d}$ . In spite of the fact that the blocks
are very sparse  we were unable to calculate the determinant of 
${\boldsymbol M}\left(l\right) $ in less than $Mm^{2}=M^{3-2/d}$ operations. 
This is only marginally better than the standard $M^{3}$ scaling. 
Furthermore in $3d$ the resulting algorithm has a very unfavorable prefactor, 
which makes this approach unpractical unless substantially improved for 
instance by better exploiting the sparsity of ${\boldsymbol M}\left( l\right) $.  

A more favorable scaling can be obtained if one focuses on the response of 
$\Omega _{f}$ with respect to appropriate parameters, which is a standard way
of calculating physical quantities. For instance the number of particles is
given by $\left\langle N\right\rangle =-\frac{\partial \Omega _{f}}{\partial
\mu }$, the energy by $\left\langle E\right\rangle =\frac{\partial (\beta
\Omega _{f})}{\partial \beta }+\mu \left\langle N\right\rangle $ and so on.
In general one can write for the  value of a property $A$, conjugated to the
field $\lambda _{A}$: 
\begin{equation}
\left \langle A\right\rangle 
=-\frac{1}{\beta }\sum_{l=1}^{P}\,\mathrm{Tr}
\left( {\boldsymbol M}(l)^{-1}
\frac{\partial {\boldsymbol M}(l)}{\partial\lambda _{A}}\right) \;.  
\label{eq:av}
\end{equation}
which requires the inversion of  the sparse matrices ${\boldsymbol M}(l)$ and
not the calculation of its  determinant. The inverse of 
${\boldsymbol M}(l)$ can be found if one solves the 
$M$ sets of linear equations 
${\boldsymbol M}(l){\boldsymbol\phi }_{j}^{l}={\boldsymbol\psi }_{j}$ where 
$\{{\boldsymbol\psi }_{j}\}$ is a complete orthonormal basis set  and is given 
by ${\boldsymbol M}(l)^{-1}=\sum_{j=1}^{M}{\boldsymbol\phi }_{j}^{l}{\boldsymbol
\psi }_{j}^{\dagger }$. Using a preconditioned biconjugate gradient 
method\cite{num_rec} and the sparsity of ${\boldsymbol M}(l)$ 
we find that solving each linear
equation takes $O\left( M\right) $ operations leading to an overall
quadratic scaling also in $3d$. An efficient preconditioner has proved to be
the inverse of ${\boldsymbol M}_{f}(l)$ for free particles. Although in this
case the full inverse can be evaluated exactly using Fourier transforms
methods it is computationally  expedient to truncate ${\boldsymbol M}%
_{f}(l)^{-1}$ so as to  give to ${\boldsymbol M}_{f}(l)^{-1}$ the same
sparse structure as ${\boldsymbol M}(l)$. As we shall see below the
theoretical  $O\left( M^{2}\right) $ scaling can be demonstrated in
practice, unfortunately at the cost of a large prefactor.

However linear scaling can be obtained if we use a stochastic approach.
Taking our cue from what is done in quantum chromodynamics
we introduce a random vector ${\boldsymbol\psi }$. 
If the $\psi _{i}$ are stochastically 
distributed such that their average satisfies: 
\begin{equation}
\left\langle \psi _{i}\psi _{j}^{\ast }\right\rangle =\delta _{ij}\quad 
\mathrm{and}\quad \left\langle \psi _{i}\right\rangle =0\;,  \label{eq:noise}
\end{equation}
the inverse of $M$ can be written as an expectation value: 
\begin{equation}
{\boldsymbol M}(l)^{-1}=\left\langle {\boldsymbol\phi }^{l}{\boldsymbol\psi }%
^{\dagger }\right\rangle \;.  \label{eq:stoc_inv}
\end{equation}
where the average is taken over the  stochastic process and 
${\boldsymbol\phi }^{l}$ is the solution of the linear equation
${\boldsymbol M}(l){\boldsymbol\phi }^{l}={\boldsymbol\psi }$. 
In principle any distribution that fulfills Eq.~(\ref{eq:noise}) 
allows finding the inverse of ${\boldsymbol M}(l)$ as in Eq.~(\ref{eq:stoc_inv}). 
However in practice it has been noted that the statistical error does 
depend on the choice of the distribution.  
The one that gave the smaller noise was the  $S_{1}$
distribution, in which the $\psi _{i}$ are distributed as $\delta (\psi
_{i}^{\ast }\psi _{j}-1)$ \cite{sha94}.  This means that the variables $\psi
_{i}$ can take random values $e^{i\alpha }$ on the unit circle with equal
probability. Physical quantities can then be calculated according to Eq.~(%
\ref{eq:av}) as: 
\begin{equation}
\left\langle Q_{\lambda }\right\rangle =-\frac{1}{\beta }\sum_{l=1}^{P}\,%
\mathrm{Tr}\left( \left\langle {\boldsymbol\phi }^{l}{\boldsymbol\psi }%
^{\dagger }\right\rangle \frac{\partial {\boldsymbol M}(l)}{\partial \lambda 
}\right) \;
\end{equation}%
and therefore one finds overall linear scaling behavior. 

We now substantiate the claims made on the scaling of the 
different algorithms introduced here with numerical calculations. 
Different model potentials have been
investigated with satisfactory results. Here we report only one
calculation done on a periodic potential constructed with a superposition of
Gaussians $\sum_{I=1}^{N}-we^{-(r-R_{I})^{2}/\delta ^{2}}\Theta \left(
r_{c}^{2}-(r-R_{I})^{2}\right) $ where $w=4\mathrm{\;a.u.}$ and the cutoff
radius is $r_{c}=6\mathrm{\;a.u.}$. The Gaussian centers $R_{I}$ are
arranged to form a cubic lattice and mimic a crystal of $N$ atoms. The
spacing between the atomic sites is taken to be $4\delta $ with $\delta =0.75%
\mathrm{\;a.u.}$ . Periodic boundary conditions are imposed throughout. The
Trotter number is chosen to be $P=256$ leading to an electronic temperature $%
T=7529K$. This is rather small on the electronic energy scale and we have
explicitly verified that it is close to the $T\rightarrow 0$ limit for the
model.

\begin{figure}[tbp]
\includegraphics*[width=8.5cm]{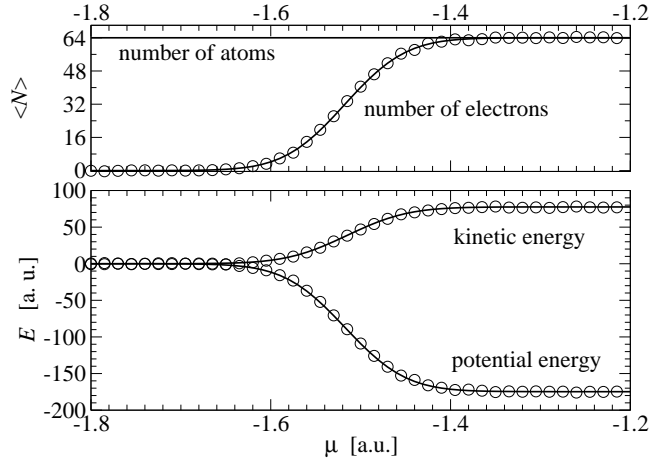}
\caption{Particle number, kinetic and potential energy of the Gaussian model
as a function of the chemical potential. The
circles show the results calculated with the stochastic linear scaling
algorithm. The lines are obtained from the smallest eigenvalues calculated
with an iterative diagonalisation algorithm.}
\label{fig:model}
\end{figure}

The number of electrons, the kinetic and the potential energy as a function
of the chemical potential are shown in Figure~\ref{fig:model}. 
Upon increasing $\mu $ the states are filled with electrons. 
At $\mu \approx-1.5\mathrm{\;a.u.}$ half of the states belonging to the first band are
occupied and the system is metallic. For $\mu \approx -1.2\mathrm{\;a.u.}$
the first band is completely filled and the model behaves as a large band
gap insulator. The number of electrons, the kinetic and the potential energy
are calculated with the stochastic algorithm and an iterative routine which
calculates the largest relevant eigenvalues for 
comparison\cite{NAG,par,rut69,rut70}. The agreement is excellent. 
In order to reach the required precision, averages needed to be
taken over about $100$ independent configurations.

In a stochastic evaluation it is important to keep track of the error which
for each property $A$ is given by $\sigma _{A}/\sqrt{N_{MC}}$
where $\sigma _{A}$ is its mean square fluctuation and $N_{MC}$ the
number of Monte Carlo steps. A number of physical observables together with
their estimated $\sigma _{A}$ is given in Table~I. It is seen that
the variances for the different observables do not differ qualitatively for
metals and insulators. Thus the number of Monte Carlo steps does not have to be
larger for metallic systems than for insulators. We have to mention that if $%
l$ is close to $P/2$ the condition number of $M(l)$ can be big in the
metallic case. This could in principle increase the number of iterations
needed to solve ${\boldsymbol M}(l){\boldsymbol\phi }_{j}^{l}={\boldsymbol\psi }_{j}$. 
In practice however this did not lead to significant problems 
and the performance of the 
algorithm in the half-filled and filled case are very similar.
For fixed Trotter number $P$ the algorithm scales linearly with 
the number of grid points as the lattice constant $\delta$ is reduced.
\begin{table}[tbp]
\begin{tabular}{rrrrr}
       & \multicolumn{2}{c}{ insulator}          & \multicolumn{2}{c}{ metal}   \\ 
       &  $ \langle A \rangle $        &   $\sigma_A$  &      $\langle A \rangle$    & $\sigma_A$ \\
$N$    & $64.0$   &  $ 2.7$  & $ 32.0 $ & $ 4.5$   \\ 
$T$    & $77.2$   &  $ 5.4$  & $ 36.7 $ & $ 4.1$   \\ 
$V$    & $-174.7$ &  $ 5.0$  & $-86.1 $ & $ 4.2$   \\ 
&  & 
\end{tabular}%
\caption{Average values and variances for the particle number $N$, the
kinetic energy $T$, the potential energy $V$ and the density at an atom site 
$\protect\rho $.}
\end{table}
\begin{figure}[tbp]
\includegraphics*[width=8.5cm]{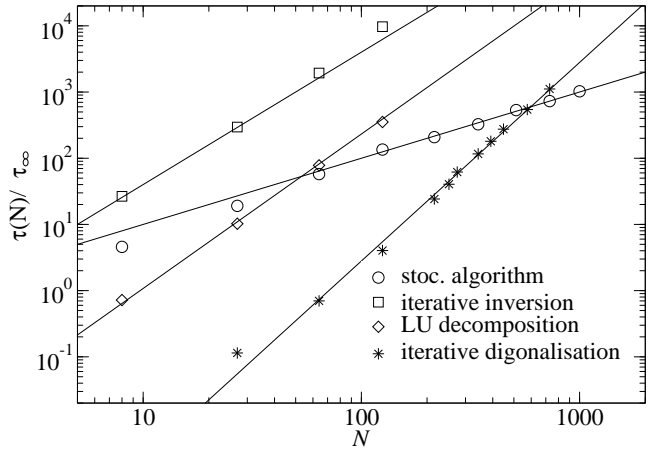}
\caption{
Log-log plot of the CPU time $\tau(N)$ versus the number of atoms N. 
The measured slopes confirm that the scaling is $O(N)$, $O(N^2)$, 
$O(N^{7/3})$ and $O(N^3)$ for the stochastic algorithm, 
the iterative inversion, the banded LU decomposition and the 
partial iterative diagonalization respectively. 
The unit of time is $\tau_\infty = \tau(N) / N$ measured at large N. 
That is the asymptotic incremental cost of an extra atom in 
the $N \rightarrow \infty $ limit. 
The computation was performed on a 1.7 GHz Pentium4 Xenon processor.}
\label{fig:scaling}
\end{figure}

We now compare the scaling of the different algorithms introduced here with
a standard diagonalization procedure in which the largest relevant
eigenvalues of $e^{\frac{\beta }{P}\left( \mu -H\right) }$ are computed with
sparse matrix diagonalisation techniques\cite{NAG,par,rut69,rut70}. 
In Figure~\ref{fig:scaling} we see that the predictions made
on the scaling of the different algorithms as a function of system size are
confirmed by actual calculations.

Comparing the performance of the stochastic method with numerically exact
methods is not easy since the performance  of the method depends on the
accuracy required. Here, in comparing the performance for different system
sizes we have instead kept the number of Monte Carlo steps constant. Had we
kept the relative accuracy constant this would have led to sublinear
scaling due to the self-averaging properties of the larger systems. With this
caveat from Figure~\ref{fig:scaling} one can see that our new algorithm
based on stochastic matrix inversion techniques scales linearly, 
while the algorithm based on matrix diagonalisation shows a cubic scaling. 
For the model chosen the crossing point at which our method becomes more 
efficient is $M\approx 38400$, which corresponds to $600$ atoms. 
It must be stressed that we have taken the worst case scenario
since we are here dealing with metals and $P$ has to be taken larger 
to reach the $T=0$ limit. At full filling convergence is reached almost for $P=128$. 
For this $P$ value the crossing point occurs at $N=450$ atoms. 
For different systems this value can vary because it might be necessary 
to choose a larger value for the Trotter number or the acceptable statistical 
error should be smaller. Still our method can be expected to become more 
efficient than cubic scaling algorithms at system sizes of a few hundred atoms. 
It should be also mentioned that the algorithm is trivially parallelizable.
Furthermore increasing the number of electrons while keeping the volume
constant does not increase the computational cost of our stochastic approach.
In contrast the deterministic methods at constant volume scale quadratically
with the number of electrons. Another advantage of the present method is
memory saving, which grows linearly with volume and not as the product of the
number of electrons times the volume. 

In order to apply this method to fully self consistent DFT 
calculation one must take into account the fact that the
evaluation of the electron density is affected by a statistical 
error.
This will be considered at a later stage. As it is now the method
can be profitably applied to tight binding calculations and to
the Harris functional approximation\cite{harris} to DFT.
It can also be extended to include the ionic positions in the sampling 
so as to obtain an efficient Monte Carlo method.

\end{document}